# Thermophysical Properties of Lignocellulose: A Cell-scale Study down to 41K

Zhe Cheng[1], Zaoli Xu[1], Lei Zhang[2], Xinwei Wang[1,*]

[1]Department of Mechanical Engineering, 2010 Black Engineering Building,

Iowa State University, Ames, IA 50011, USA

[2]Department of Foreign Language, Hubei University of Technology,

Wuhan, Hubei 430065, P. R. China

**Abstract**

Thermal energy transport is of great importance in lignocellulose pyrolysis for biofuels. The thermophysical properties of lignocellulose significantly affect the overall properties of bio-composites and the related thermal transport. In this work, cell-scale lignocellulose (mono-layer plant cells) is prepared to characterize their thermal properties from room temperature down to ~40 K. The thermal conductivities of cell-scale lignocellulose along different directions show a little anisotropy due to the cell structure anisotropy. It is found that with temperature going down, the volumetric specific heat of the lignocellulose shows a slower decreasing trend against temperature than microcrystalline cellulose, and its value is always higher than that of microcrystalline cellulose. The thermal conductivity of lignocellulose decreases with temperature from 243 K to 317 K due to increasing phonon-phonon scatterings. From 41 K to 243 K, the thermal conductivity rises with temperature and its change mainly depends on the heat capacity's change.

---

[*] Corresponding author. Email: xwang3@iastate.edu, Tel: 001-515-294-8023.

# 1. Introduction

Lignocellulose in nature is the most abundant raw material from wood, grasses, and agricultural residues.[1] It not only can be used as combustion fuels directly, also can be converted to various forms of biofuels indirectly through pyrolysis or hydrolysis.[2-5] Moreover, it has been extensively reported as filler to make lignocellulose-based biodegradable composites [6, 7] and thermoplastic composites.[8, 9]

Mettler *et al.* listed ten top fundamental challenges of biomass pyrolysis for biofuels and a major challenge of them is the lack of accurate knowledge of heat and mass transfer properties (*e.g.*, thermal conductivity, diffusivity).[5] So the thermal properties of lignocellulose are in high demand, especially the thermal properties at the cell scale. This is because microstructure provides more details and fundamental information about lignocellulose compared with the bulk material that only provides an overall value of thermophysical properties. Also, for bio-composites, thermophysical properties are those of the most important parameters. The thermophysical properties of filled material (lignocellulose) determine the overall thermal property of the composite together with the matrix material. Therefore, it is of great importance to study the thermal transport in lignocellulose at the cell scale. Such information provides the critical knowledge base for evaluating the overall thermophysical properties of bio-composites.

Due to the importance of this topic, many researchers studied the bulk thermal properties and their anisotropy of various forms of lignocellulose, including wood, jute, cotton, sisal and ramie. Specifically, Yapici *et al.* measured the thermal conductivity of beech, oak, fir, scots pine and chestnut. They found the thermal conductivity of them ranged from 0.18 to 0.40 $Wm^{-1}K^{-1}$ and the thermal conductivity parallel to the grain angle was a little larger than that perpendicular to the grain angle.[10] Gupta *et al.* reported the thermal conductivity of softwood, softwood bark and softwood char at 310 K as 0.1 $Wm^{-1}K^{-1}$, 0.2 $Wm^{-1}K^{-1}$ and 0.1 $Wm^{-1}K^{-1}$ respectively.[11] Alsina *et al.* studied the thermal conductivity of hybrid lignocellulosic fabrics and obtained the thermal conductivity of jute/cotton, sisal/cotton and ramie/cotton hybrid fabrics as 0.185 $Wm^{-1}K^{-1}$, 0.575 $Wm^{-1}K^{-1}$ and 0.555 $Wm^{-1}K^{-1}$ when the heat transfer was parallel to the fabrics and 0.19 $Wm^{-1}K^{-1}$, 0.415 $Wm^{-1}K^{-1}$, 0.36 $Wm^{-1}K^{-1}$ when the heat flux was perpendicular to the fabrics.[12] Stankovic *et al.* measured the thermal conductivity of cotton, hemp and Hemp/cotton as 0.026 $Wm^{-1}K^{-1}$, 0.022 $Wm^{-1}K^{-1}$ and 0.034 $Wm^{-1}K^{-1}$. They attributed the low thermal conductivity to porous structures of these fibers.[13] In Incropera's book, the thermal conductivity of cotton is 0.06 $Wm^{-1}K^{-1}$.[14] The variation in thermal conductivity of cotton is due to sample-to-sample difference.

The thermal conductivity of bulk lignocellulose has been extensively studied. However, thermal transport in lignocellulose at the cell scale (mono-layer plant cells) has not been studied before. Even for all other cells, very few measurements have been conducted on thermal properties at the cellular level due to great challenges like sample's small size and difficulty in sample handling and thermal probing. To our best knowledge, only Park *et al.* measured the thermal conductivity of single animal cells based on the three-omega method.[15]

This work reports the first effort to characterize the thermal transport in lignocellulose at the cell scale. To overcome experimental difficulties like small size and difficult manipulation during thermal transport study, inner epidermis of onion bulb scale is chosen in this work because it is composed of mono-layer plant cells.[16] Thermal transport capacity measurement in lignocellulose along single cells can be achieved by characterizing the thermal conductivity of onion inner epidermis membrane. Moreover, the thermal transport capacity along different directions can be evaluated to explore the anisotropic thermal conductivity of lignocellulose at the cell scale. Our result provides pioneering insight into micro/nanoscale scale heat transfer mechanism of biomass pyrolysis and bio-composite. Moreover, onion inner epidermis membrane has been extensively reported to use as soft support of crystal grown, biosensors and enzyme immobilization.[17-19] Temperature change in the soft substrate would affect these processes significantly. Therefore, it is of great importance to know the substrate's thermal properties.

In this work, we use the transient electro-thermal (TET) technique to characterizing the thermal transport in cell-scale lignocellulose under temperatures ranging from room temperature down to 20 K. Four samples of cell-scale lignocellulose are prepared from one piece of inner epidermis of onion bulb scale for various purposes. Two samples with different lengths are studied to evaluate the samples' emissivity, which is used to substrate the radiation effect in the measurement. Another two samples are used to explore the anisotropy of energy transport capacity along different directions. In addition, a robust method developed by our lab is employed to determine the Lorenz number of Iridium film on lignocellulose. This helps subtract the Iridium's effect on sample's thermal diffusivity. Finally, the relation between thermal properties of cell-scale lignocellulose and temperature are obtained and discussed.

## 2. Materials and Methods

### 2.1. Sample preparation

The onion strips used in this work are from inner epidermis of onion (Allium cepa) bulb scales,

which are composed of mono-layer cells. First, a small piece of onion is torn off from an onion bulb as shown in Fig. 1. Then the membrane from the inner side from this piece of onion is fixed to a glass slide after peeled off with tweezers. This keeps the membrane flat when the water in onion cells evaporates in air. Twenty four hours later, the membrane composed of mono-layer cells is separated from the glass slide, and then cut into long strips as the to-be-measured samples with the help of blade and optical microscope.

In this work, we prepare four samples with different lengths and cell directions for different purposes. Specifically, Sample 1 is a short sample and its measured energy transfer is along the cell length direction while Sample 2 is a long sample with the same measured direction. The definition of directions is shown as Fig. 1(d). In this figure, it is evident that the cells have an oval shape. Sample 3 has a similar length with Sample 1 but its measured energy transfer is along the cell width direction. Sample 4 has similar length and measured direction with Sample 1, which is prepared to determine the Lorenz number of Iridium thin film on onion cells. These samples are prepared from the same small piece of onion membrane, so they have similar properties. All their dimensions are summarized in Table 1.

As important parameters in characterizing samples' thermal diffusivity, geometries (length, width and average thickness) of these samples need to be determined. A scanning electron microscopy (SEM) is used to take pictures and measure the lengths and widths of these samples. The samples' thickness is not uniform in the cross section and cannot be measured directly. To obtain the average thickness of the onion cell, a thin glass slide is put into a tube after a sample is fixed on the glass slide. Because the glass slide's width is the same with the tube's diameter, the glass slide lies parallel to the tube's axis accurately. This guarantees that the sample also lies parallel to the tube's axis accurately. Then one end of the tube is sealed and epoxy resin is added. Then the sample and glass slide are fixed through epoxy resin's curing reaction. When the epoxy resin solidifies completely, the cylinder epoxy resin which contains the sample is polished. Then, an epoxy resin cylinder's smooth cross section can be obtained. Meanwhile, the sample's cross section can also be obtained, as shown in Fig. 2. Finally, the average thickness is calculated as 1124 nm. During thickness evaluation, the thicknesses of different sections along the cross section are measured with SEM.

To carry out TET measurement, samples must be conductive, so the onion cells are coated with Iridium using a sputtering machine Quorum Q150T S. The thicknesses of the Iridium films are monitored using a quartz crystal microbalance during deposition.

## 2.2. Thermal transport characterization

The transient electro-thermal (TET) technique is an effective, accurate, and fast approach developed in our lab to measure the thermal diffusivity of solid materials, including one-dimensional conductive, semi-conductive or non-conductive structures. The measurement accuracy and effectiveness of TET technique has been fully examined by characterizing both metallic and dielectric materials. Using the TET technique, Guo *et al.* measured the thermal diffusivity of micro-scale polyester fibers.[20] Furthermore, Feng *et al.* characterized the thermal diffusivity of thin films constituted of anatase $TiO_2$ nanofibers and free-standing micrometer-thick Poly (3-hexylthiophene) films.[21, 22] The TET measurement results showed a good agreement with reference values.

In this work, the TET technique is used for thermal characterization of cell-scale lignocellulose.

A schematic of the TET technique is presented in Fig. 3(a) to indicate how this technique is used to characterize the thermal transport in one-dimensional micro/nanostructures. At the beginning, the to-be-measured sample is suspended between two aluminum electrodes and coated with Iridium film on the top side of sample to make it electrically conductive. Silver paste is used to enhance the electrical and thermal contact between the sample and electrodes at the contact points. Then the sample is placed in a high vacuum chamber to suppress the effect of gas conduction during thermal characterization. Figures 3(b) and (c) show the SEM pictures of the mono-layer cell samples connected between two electrodes. From Fig. 3(c) we can clearly see the cells are aligned in the length direction between the electrodes while the cell-cell contacts are not clear in Fig. 3(b). This is because the cell-cell contacts show different appearance on the two surfaces of mono-layer cells. We can see from Fig. 2(a) that the cell-cell contacts in the upper side are more protruding than those in the lower one. Figure 3(c) is the picture of the upper side surface and Fig. 3(b) is the picture of the lower side one.

In the experiment, a periodic step DC current is fed through the sample to generate joule heat. The electrical current level needs to be chosen carefully to guarantee that the temperature rise of all the TET experiments is moderate while still be sensible. The temperature evolution of the sample is determined by two competing processes: one is joule heating by the electrical current, and the other one is heat conduction along the sample to electrodes. The temperature change of the sample will induce an electrical resistance change, which leads to an overall voltage change of the sample. Therefore, the voltage change of the sample can be used to monitor its temperature evolution. A typical *V-t* profile recorded by the oscilloscope for Sample 1 is shown in Fig. 3(d). More details about the TET technique can be referred to Feng's work.[22]

During TET thermal characterization, the surface radiation effect could be significant at and above room temperature if the sample has a very large aspect ratio (*L/D*, *L*, *D*: length and thickness of sample). The heat transfer rate from the sample surface due to radiation can be expressed as:

$$Q_{rad} = \varepsilon_r \sigma A_s (T^4 - T_0^4) = 2\varepsilon_r \sigma WL \left(4T_0^3 \theta + 6T_0^2 \theta^2 + 4T_0 \theta^3 + \theta^4\right), \quad (1)$$

where $\varepsilon_r$ is the effective emissivity of the sample, $\sigma = 5.67 \times 10^{-8}$ Wm$^{-2}$K$^{-4}$ is the Stefan-Boltzmann constant, $A_s$ the surface area, $W$ the width, $L$ the length, $T$ the surface temperature, $T_0$ the ambient temperature (vacuum chamber), and $\theta = T - T_0$ is the temperature rise. When $T_0$ is relatively large, $\theta \ll T_0$ then we have $Q_{rad} \approx 8\varepsilon_r \sigma WL T_0^3 \theta$. When $T_0$ is small, the surface radiation effect is negligible, and we can still use $Q_{rad} \approx 8\varepsilon_r \sigma WL T_0^3 \theta$ to estimate the surface radiation effect.

To eliminate heat convection, the sample is measured in a high vacuum chamber in which the pressure is down to 0.4-0.5 mTorr (detected by a convection vacuum gauge, CVM211 Stinger, InstruTech). This high vacuum level can make sure that the residual water in the samples evaporates easily and quickly and the effect of water content would not affect the measured properties. Ignoring the gas conduction and non-consistent heating and convection, which are negligible in our experiment, and converting the surface radiation to body cooling source, the heat transfer governing equation for the sample becomes:

$$\frac{1}{\alpha} \frac{\partial \theta(x,t)}{\partial t} = \frac{\partial^2 \theta(x,t)}{\partial x^2} + \frac{I^2 R_0}{kLA_c} + \frac{1}{k} \frac{8\varepsilon_r \sigma T_0^3}{D} \theta, \quad (2)$$

where $\alpha$ is thermal diffusivity, $k$ thermal conductivity and $A_c$ the cross-sectional area. $I$ is the electrical current passing through the sample and $R_0$ is the sample's resistance before heating.

$q_0 = I^2 R / WDL$ is the electrical heating power per unit volume and is constant during measurement. Integral of Green's function[23] is used to solve the partial differential equation (2) and the average temperature along the sample can be obtained:

$$\bar{T} = T_0 + \frac{q_0 L^2}{12} \frac{48}{\pi^4} \sum_{m=1}^{\infty} \frac{1-(-1)^m}{m^2} \frac{1-\exp\left[-(m^2-f)\pi^2(\alpha t/L^2)\right]}{(m^2-f)}. \quad (3)$$

When time goes to infinity, the temperature distribution along the sample will reach a steady state. The average temperature of the sample in the final steady state is:

$$T(t \to \infty) = T_0 + \frac{q_0 L^2}{12k}. \quad (4)$$

More details regarding the solution are provided in reference.[24] With an effective thermal diffusivity $\alpha_{eff} = \alpha \cdot (1-f)$, here $f$ is defined as $-8\varepsilon_r \sigma T_0^3 L^2 / D\pi^2 k$, the normalized average temperature rise $T^*$ is:

$$T^* \cong \frac{48}{\pi^4} \sum_{m=1}^{\infty} \frac{1-(-1)^m}{m^2} \frac{1-\exp[-m^2\pi^2 \alpha_{eff} t/L^2]}{m^2}. \quad (5)$$

The measured voltage change is inherently proportional to the temperature change of the sample. The normalized temperature rise $T^*$ is calculated from experiment as $T^* = (V_{sample} - V_0)/(V_1 - V_0)$, where $V_0$ and $V_1$ are the initial and final voltages across the sample. In our work, after $T^*$ is obtained, different trial values of $\alpha_{eff}$ are used to calculate the theoretical $T^*$ using equation (5) and fit the experimental result. The value giving the best fit of $T^*$ is taken as the effective thermal diffusivity of the sample. The determined thermal diffusivity still has the effect of parasitic conduction because the measured sample is coated with a thin Iridium film. The thermal transport effect caused by the coated layer can be subtracted using the Wiedemann-Franz law with a negligible uncertainty. The real thermal diffusivity ($\alpha_{real}$) of the sample is determined as:

$$\alpha_{real} = \alpha - \frac{L_{Lorenz}TL}{RA_c(\rho c_p)}, \tag{6}$$

where $\rho c_p$ is volume-based specific heat of the sample, which can be obtained from calibration in our cryogenic system to be discussed later. $L_{Lorenz}$, $T$, and $A_c$ are the Lorenz number, sample's average temperature and cross-sectional area, respectively. The Lorenz number can be determined from experiment discussed later. In summary, the real thermal diffusivity of the sample finally is calculated as:

$$\alpha_{real} = \alpha_{eff} - \frac{1}{\rho c_p} \frac{8\varepsilon_r \sigma T_0^3}{D} \frac{L^2}{\pi^2} - \frac{L_{Lorenz}TL}{RA(\rho c_p)}, \tag{7}$$

The thermal conductivity can be readily calculated as $k_{real}=\alpha_{real}\rho c_p$.

## 3. Results and discussion

### 3.1. The surface emissivity and Lorenz number of Ir

First of all, we take Sample 1 as an example to demonstrate how the thermal diffusivity is characterized. This sample is 1352 μm long and 527 μm wide. The energy transfer direction we measured is along the cell length. The electrical resistances before and after applying a step current are 44.82 Ω and 45.49 Ω, and the electrical current used in the experiment is 2.4 mA to give a change of voltage increase at about 1.5%. Figure 3(d) shows the transient voltage change of raw experimental data. The normalized temperature rise $T^*$ and the fitting result are shown in Fig. 3(e). Its effective thermal diffusivity is determined as $6.78 \times 10^{-7}$ m$^2$ s$^{-1}$, which includes the effect of radiation and parasitic conduction. We vary the trial values of $\alpha$ to determine the fitting uncertainty as shown in Fig. 3(e). When the trial value is changed by 10%, the fit can be seen deviating from the experimental results substantially. So the uncertainty is determined as ±10%.

In order to eliminate the effect of radiation, sample's surface emissivity $\varepsilon_r$ is needed. Experiments on two samples from the same onion piece with different sizes (Sample 1 and Sample 2) are carried out to determine the surface emissivity $\varepsilon_r$, based on

$$k_{real} = \alpha_{eff1}\rho c_p - \frac{8\varepsilon_r \sigma T_0^3}{D}\frac{L_1^2}{\pi^2} - \frac{L_{Lorenz}T_1L_1}{R_1DW_1} \text{ and } k_{real} = \alpha_{eff2}\rho c_p - \frac{8\varepsilon_r \sigma T_0^3}{D}\frac{L_2^2}{\pi^2} - \frac{L_{Lorenz}T_2L_2}{R_2DW_2},$$

where $D$ is average thickness of onion cell, subscript 1, 2 refer Sample 1 and Sample 2, respectively. $\rho c_p$ and $L_{Lorenz}$ will be determined using different techniques and mentioned latter. We use the volumetric specific heat of Sample 1 [$1.80 \times 10^6$ J K$^{-1}$ m$^{-3}$)] to do the calculation and the Lorenz number is $6.34 \times 10^{-9}$ W·Ω·K$^{-2}$. The length and width of Sample 2 are 1996 μm and 458 μm respectively. During the TET characterization, its electrical resistance rises from 72.6 Ω to 73.6 Ω. The effective thermal diffusivity of Sample 1 and Sample 2 are $6.78 \times 10^{-7}$ m$^2$ s$^{-1}$ and $9.43 \times 10^{-7}$ m$^2$ s$^{-1}$ respectively. Only $k_{real}$ and $\varepsilon_r$ are unknown in the two equations above. The surface emissivity is determined as 0.22 by solving the two equations.

To measure metallic film's Lorenz number, a method involving repeatedly depositing metallic films is developed in our lab and Sample 4 is used to determine the Lorenz number of the Iridium film on onion cells. It can be seen from equation (7) that $\alpha_{eff}$ is proportional to $R^{-1}$. Iridium film layers with various thicknesses on Sample 4 would have different effective thermal diffusivity $\alpha_{eff}$ and electrical conductance $R^{-1}$. Thus, a 5 nm-thick iridium film layer is deposited repeatedly on Sample 4 for five times. Accordingly, five effective thermal diffusivity values and five $R^{-1}$ values are obtained. These data can be fitted into a straight line and its slope is $L_{Lorenz}TL/A_0(\rho c_p)_0$ (the subscript "0" represents onion mono-layer cells' parameters). More details can be found in the reference.[25] The Lorenz number of Iridium film on Sample 4 is determined as $6.34 \times 10^{-9}$ W·Ω·K$^{-2}$. This value is much smaller than that of bulk material ($2.49 \times 10^{-8}$ W·Ω·K$^{-2}$), which results from many factors, including Iridium film's non-uniform nanostructure due to rough cell wall surface and cell-to-cell contacts, defects and gain boundaries in the metallic film.

## 3.2. Anisotropy of thermal transport

The epidermal cell wall consists of layers of parallel, cellulose micro fibrils and its structure is like 'plywood laminates', which contain micro fibrils orientated in all directions.[26] The onion epidermis cell wall can be seen as fiber-reinforced composite materials,[27] and onion cellulose fibrils have an overall orientation parallel with the cell length.[16] Due to anisotropy of onion cells' strip structure and cellulose orientation, energy transport along different directions may be different. Thus, the thermal properties of Sample 1 (energy transport direction is along the cell length) and Sample 3 (energy transport direction is along the cell width) are measured at ambient temperature $T_0$=290 K and are compared to study the anisotropic nature. The thermal conductivity of Sample 1 is 0.72 Wm$^{-1}$K$^{-1}$ while that of Sample 3 is 0.81 Wm$^{-1}$K$^{-1}$. This indicates that thermal transport capacity along the cell width is better than that along the cell length although this difference is quite small. This is in contrary to most of the bulk material measured in the literature.[10, 12] This could be due to onion cells' strip structure. Obviously, energy transport along the cell width has to travel through more cell-to-cell contacts per unit length than that along the cell length direction. These cell-to-cell contacts are thicker and more condensed than other parts of cells, which would promote more energy transport. Therefore, the thermal conductivity along the cell width direction is larger than that along the cell length direction. This thermal conductivity difference indicates energy transport through these strip cells shows a little bit anisotropy.

### 3.3. Effect of temperature on thermal transport

Sample 2 is used to characterize energy transport in cell-scale lignocellulose from room temperature down to 20 K in a cryogenic system (CCS-450, JANIS). The result of temperature dependent electrical resistance and electrical resistivity of Sample 2 is displayed in Fig. 4. It can be seen that both the electrical resistance and resistivity decrease from 72.61 Ω to 60.95 Ω when the ambient temperature falls from 290 K to 20 K. A linear approximation is typically used to describe the relation between electrical resistance and temperature: $R = R_0[1+\eta(T-T_0)]$ when temperature is not too low. Here $T_0$ (290 K) is room temperature, $R_0$ is the electrical resistance at room temperature $T_0$ and $\eta$ is the temperature coefficient of resistance, determined as $6.10 \times 10^{-4}$ $K^{-1}$ in our work. After we finish the cooling process, we allow the sample's temperature slowly rise back to the room temperature. We found that at room temperature, the electrical resistance becomes 70.4 Ω, which is 2.2 Ω smaller than that before cryogenic experiment. This is caused by permanent structure change of the sample due to thermal shrink when the temperature is decreased down to 20 K. We assume this change is linear with temperature. Accordingly, $\eta$ is determined as $4.98 \times 10^{-4}$ $K^{-1}$ after subtracting the effect of sample's structure change. With knowledge of $\eta$, the temperature rise during our TET characterization can be obtained from electrical resistance change (ΔR) as $\Delta T = \Delta R/(\eta R_0)$. The slope of electrical resistivity change with temperature $d\rho_{resis}/dT$ is determined as $1.47 \times 10^{-10}$ $\Omega m K^{-1}$, which is smaller than that of bulk Iridium ($2.12 \times 10^{-10}$ $\Omega m K^{-1}$). According to Matthiessen's Rule, this slope of nanofilm and bulk material should be close. In our laboratory, Iridium nanofilms on several different substrates have been measured. This value of Iridium film on milkweed floss fiber is $1.52 \times 10^{-10}$ $\Omega m K^{-1}$ and on glass fiber is $1\sim1.5 \times 10^{-10}$ $\Omega m K^{-1}$. Only the values of film on DNA fiber and silkworm silk are the same as that of bulk Iridium. Further research is under process in our laboratory to explore the mechanisms of the resistivity temperature coefficient and the effect of the substrate.

From equation (4) $\Delta T = q_0 L^2/(12 k_{eff})$, an effective thermal conductivity can be obtained as $k_{eff} = q_0 L^2/(12\Delta T)$. Meanwhile the effective thermal diffusivity is determined by the TET measurements. Therefore, the volumetric specific heat at various temperatures can be determined by $\rho c_P = k_{eff}/\alpha_{eff}$. The results of $\rho c_P$ are shown in Fig. 5. From this figure, we can see that $\rho c_P$ increases almost linearly against temperature. Lattice vibrations make dominant contributions to lignocellulose's specific heat because the contribution of thin metallic film to the whole specific heat is negligible. At room temperature, the lignocellulose' volumetric specific heat is near the counterpart of microcrystalline cellulose and this value decreases with temperature. The reason is that short wavelength phonons are frozen out at low temperature and only long wavelength phonons contribute to specific heat, resulting in the decreasing specific heat. Meanwhile density changes little with temperature when compared with heat capacity. Therefore, the volumetric specific heat decreases with temperature. Blokhin *et al.* obtained the relation between heat capacity of plant microcrystalline cellulose and temperature via experiments.[28] The density of cellulose is 1500 Kg m$^{-3}$. The volumetric specific heat is determined and shown in Fig.5 for comparison. Their heat capacity decreases with temperature linearly and has the same trend with the results of cell-scale lignocellulose. But when temperature goes lower than room temperature, the cell-scale lignocellulose's volumetric specific heat is larger than that of cellulose microcrystals. This is due to the structure difference of cell-scale lignocellulose and pure cellulose microcrystals.

The cell-scale lignocellulose (onion cells) is composed of cellulose, hemicellulose and lignin. It contains a large amount of cellulose micro fibrils in onion epidermis cell walls.[16, 26, 27, 29] The structure of cell-scale lignocellulose is amorphous according to the X-Ray Diffraction results which will be discussed in detail later while cellulose fibers in the literature are microcrystalline. It is well documented that the specific heat can be separated into harmonic and anharmonic terms and the anharmonic term is volume and explicit temperature dependent.[30, 31] The volume of amorphous state expands when compared with that of the crystalline state due to the disordered and irregular structure in the amorphous state. So the increased volume results in an enhanced anharmonic contribution to the specific heat. Furthermore, the excessive volume reduces the interaction energy of atoms, results in weaker interatomic bonding, subsequently a reduced Debye temperature and enhanced specific heat. Additionally, the impurities and thermal expansion would also affect the specific heat of the sample. The cell-scale lignocellulose is composed of not only cellulose, but also hemicellulose, pectin and inside materials. So the specific heat of these materials would affect the overall specific heat. Also, the density used to calculate the volumetric specific heat of microcrystal cellulose fibers is the one at room temperature. This value should be a little larger due to thermal shrink when the temperature goes down, which means that the volumetric specific heat of microcrystal cellulose fiber should be a little larger. Overall, the higher excess specific heat results from the greater irregularity and disorder or the higher impurity contamination in the structure.[32, 33]

Alsina *et al.* studied the volumetric specific heat of hybrid lignocellulosic fabrics and obtained the volumetric specific heat of jute/cotton, sisal/cotton and ramie/cotton hybrid fabrics as $1.04 \times 10^6$ J m$^{-3}$K$^{-1}$, $1.13 \times 10^6$ J m$^{-3}$K$^{-1}$, $1.07 \times 10^6$ J m$^{-3}$K$^{-1}$ when the heat flux is parallel to the fabrics.[12] Gupta *et al.* measured the specific heat and density of softwood, softwood bark and softwood chars. The volumetric specific heat which can be determined as the product of specific heat and density of them is $4.22 \times 10^5$ J m$^{-3}$K$^{-1}$, $6.57 \times 10^5$ J m$^{-3}$K$^{-1}$ and $2.30 \times 10^5$ J m$^{-3}$K$^{-1}$ respectively.[11] These bulk values are much smaller than those of cell-scale lignocellulose and cellulose microcrystals and it is due to their porous structure.[13]

According to equation (7), the real thermal conductivity of cell-scale lignocellulose at different temperatures can be determined after subtracting the effect of Iridium film's parasitic conduction and radiation. Moreover, to explore the structure of cell-scale lignocellulose, X-Ray Diffraction (XRD) is used in this work. The XRD system (Siemens D 500 diffractometer) is equipped with a copper tube that was operated at 45kV and 30mA. The sample was placed on a zero-background holder (ZBH) for analysis. The zero-background holder was also scanned without the sample to provide a "blank" diffractogram. The sample was scanned with two-theta from 3 to 70 degrees using a 0.05 degree step and with a dwell time of 3 seconds per step. The thermal conductivity results are shown in Fig. 6 and the XRD results are depicted in the inset, which shows that the cell-scale lignocellulose is amorphous. It can be seen from the figure that there exists a peak value: 0.97 Wm$^{-1}$K$^{-1}$ at 243 K. When the temperature is larger than this value, the cell-scale lignocellulose's thermal conductivity decreases with increasing temperature. This is due to phonon-phonon scattering, which is dominant at high temperatures and is stronger when the temperature goes up. Phonon-phonon scattering intensifies with increasing temperature, resulting in reduced phonon mean free path and accordingly reduced thermal conductivity. When temperature is lower than 243 K, the cell-scale lignocellulose's thermal conductivity decreases with decreasing temperature. In this temperature range, defects, impurity and boundary scattering become the dominant effects which limit phonon mean free path. In this case, the change of thermal conductivity mainly depends on the change of sample's specific heat, while the phonon mean free path (mainly determined by defects, impurity and boundary scattering) has little change with temperature. The measured thermal conductivity is that of an amorphous structure. For crystalline cellulose, the thermal conductivity should be much higher. No report is available for the thermal conductivity of crystalline cellulose for comparison. Crystalline cellulose nanofibers were reported as high thermoconductive phase to produce cellulose nanofiber/epoxy resin nanocomposite. An overall thermal conductivity of the nanocomposite material was achieved over 1 $Wm^{-1}K^{-1}$.[34] It is expected the thermal conductivity of crystalline cellulose should be much higher than 1 $Wm^{-1}K^{-1}$.

The thermal conductivity of cell-scale lignocellulose is 0.73 $Wm^{-1}K^{-1}$ at 317 K while those of hard wood and soft wood are 0.16 $Wm^{-1}K^{-1}$ and 0.12 $Wm^{-1}K^{-1}$ respectively.[14] The thermal conductivity of wood is much lower than that of mono-layer cells because wood is of high porosity and full of air. Air has a very low thermal conductivity ($2.63\times10^{-2}$ $Wm^{-1}K^{-1}$), which reduces the overall thermal conductivity of wood. The samples used in this work are well-aligned and solid, which gives a relatively high thermal diffusivity and density, and accordingly a large thermal conductivity. Sakuratani measured the thermal conductivity of rice stem as 0.54 $Wm^{-1}K^{-1}$,[35] and the thermal conductivity of sisal/cotton hybrid fabrics is 0.575 $Wm^{-1}K^{-1}$ when the heat flux was parallel to the fabrics.[12] They are close to, but still smaller than the thermal conductivity of cell-scale lignocellulose. The thermal conductivity of mono-layer cells at 317 K is 0.73 $Wm^{-1}K^{-1}$ while the thermal conductivity of water is 0.63 $Wm^{-1}K^{-1}$ at this temperature. When temperature decreases down to 273 K, the thermal conductivity of mono-layer cells increases to 0.87 Wm$^{-1}$K$^{-1}$ while that of water decreases to 0.56 Wm$^{-1}$K$^{-1}$.[36] This may help plant cells keep a relative stable overall thermal conductivity when the environment temperature fluctuates around room temperature, which can facilitate the stability of live organisms.

## 4. Conclusions

This work reported a detailed study of the thermophysical properties of lignocellulose at the cell scale. Around room temperature, $\rho \cdot c_p$ of the mono-layer cells is close to that of microcrystalline cellulose. With temperature going down, $\rho \cdot c_p$ of the mono-layer cells shows a slower decrease against temperature than microcrystalline cellulose, and is always higher than that of microcrystalline cellulose. The thermal conductivity of cell-scale lignocellulose decreased with increasing temperature from 243 K to 317 K due to increasing phonon-phonon scatterings. From 41 K to 243 K, the thermal conductivity went up with temperature, mainly due to the change in heat capacity.

**Acknowledgement**

Support of this work by Army Research Office (W911NF-12-1-0272), Office of Naval Research (N000141210603), and National Science Foundation (CBET1235852, CMMI1264399) is gratefully acknowledged. X.W thanks the partial support of the "Eastern Scholar" Program of Shanghai, China.

**List of Tables and Figures**

Table 1  Dimension of samples measured in this work.

Figure 1  Sample preparation process. (a) Onion bulb. (b) Bulb scale. (c) Inner epidermis of onion bulb scale (cell-scale lignocellulose, mono-layer plant cells). (d) Magnified cell structure and direction definition.

Figure 2  Cross section of cell-scale lignocellulose and the thickness measurement: (a) cross section of sample. (b-d) magnified figure to show the thickness in different sections.

Figure 3  (a) Schematic of the experimental principle of the TET technique to characterize the thermal diffusivity of cell-scale lignocellulose. (b-c) SEM graphs of a sample connected between two electrodes. From figure (c) we can clearly see the cell is connected between the two electrodes in the cell length direction. (d) A typical *V-t* profile recorded by the oscilloscope for Sample 1 induced by the step DC current. (e) TET fitting results for mono-layer cells (Sample 1). The figure consists of the normalized experimental temperature rise versus time, theoretical fitting results, and another two fitting curves with ±10% variation of $\alpha_{eff}$ to demonstrate the uncertainty of the fitting process (blue line is for +10%, and green one is for -10%).

Figure 4  Temperature dependence of electrical resistance and resistivity of Sample 2.

Figure 5  Temperature dependence of the cell-scale lignocellulose's volumetric specific heat.

Figure 6  Temperature dependence of cell-scale lignocellulose's thermal conductivity. The inset shows the X-Ray Diffraction results of cell-scale lignocellulose.

**Table 1. Dimension of samples measured in this work**

| Sample index | Length (μm) | Width (μm) | Direction of thermal transport |
|---|---|---|---|
| 1 | 1352 | 527 | cell length |
| 2 | 1996 | 458 | cell length |
| 3 | 1082 | 428 | cell width |
| 4 | 942 | 181 | cell length |

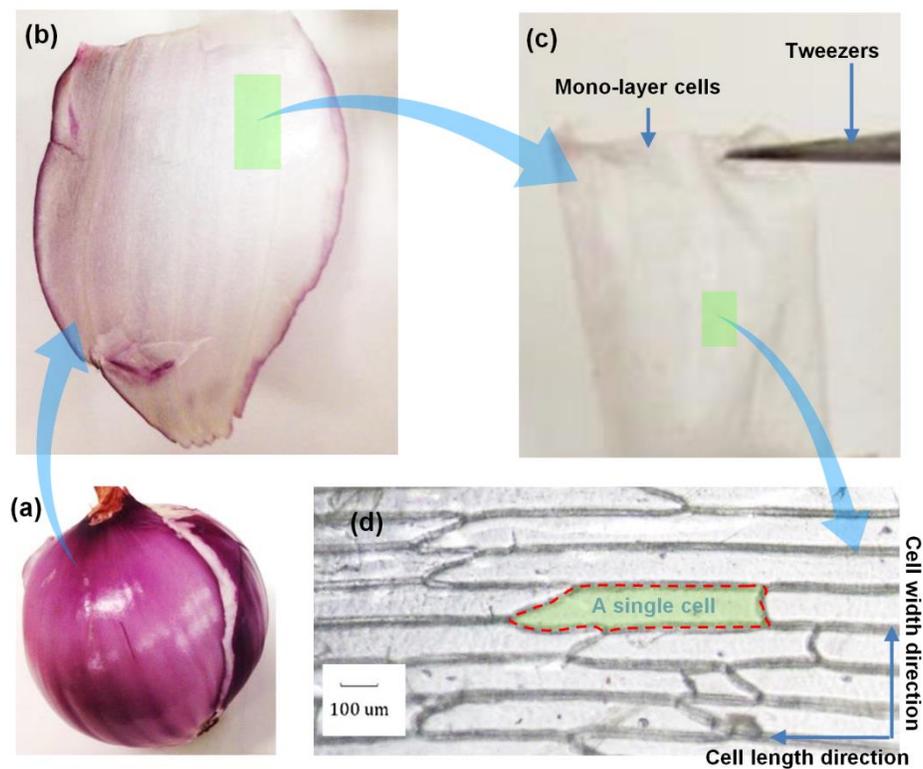

**Fig. 1** Sample preparation process. (a) Onion bulb. (b) Bulb scale. (c) Inner epidermis of onion bulb scale (cell-scale lignocellulose, mono-layer plant cells). (d) Magnified cell structure and direction definition.

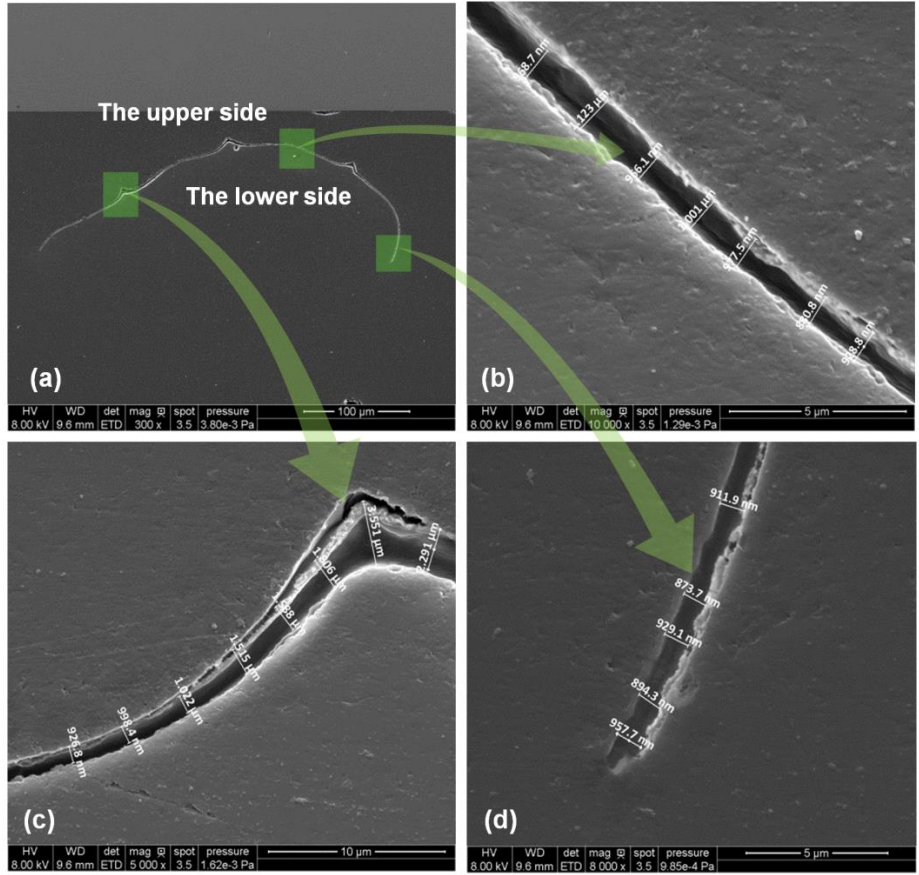

**Fig. 2** Cross section of cell-scale lignocellulose and the thickness measurement: (a) cross section of a sample. (b-d) magnified figure to show the thickness in different sections.

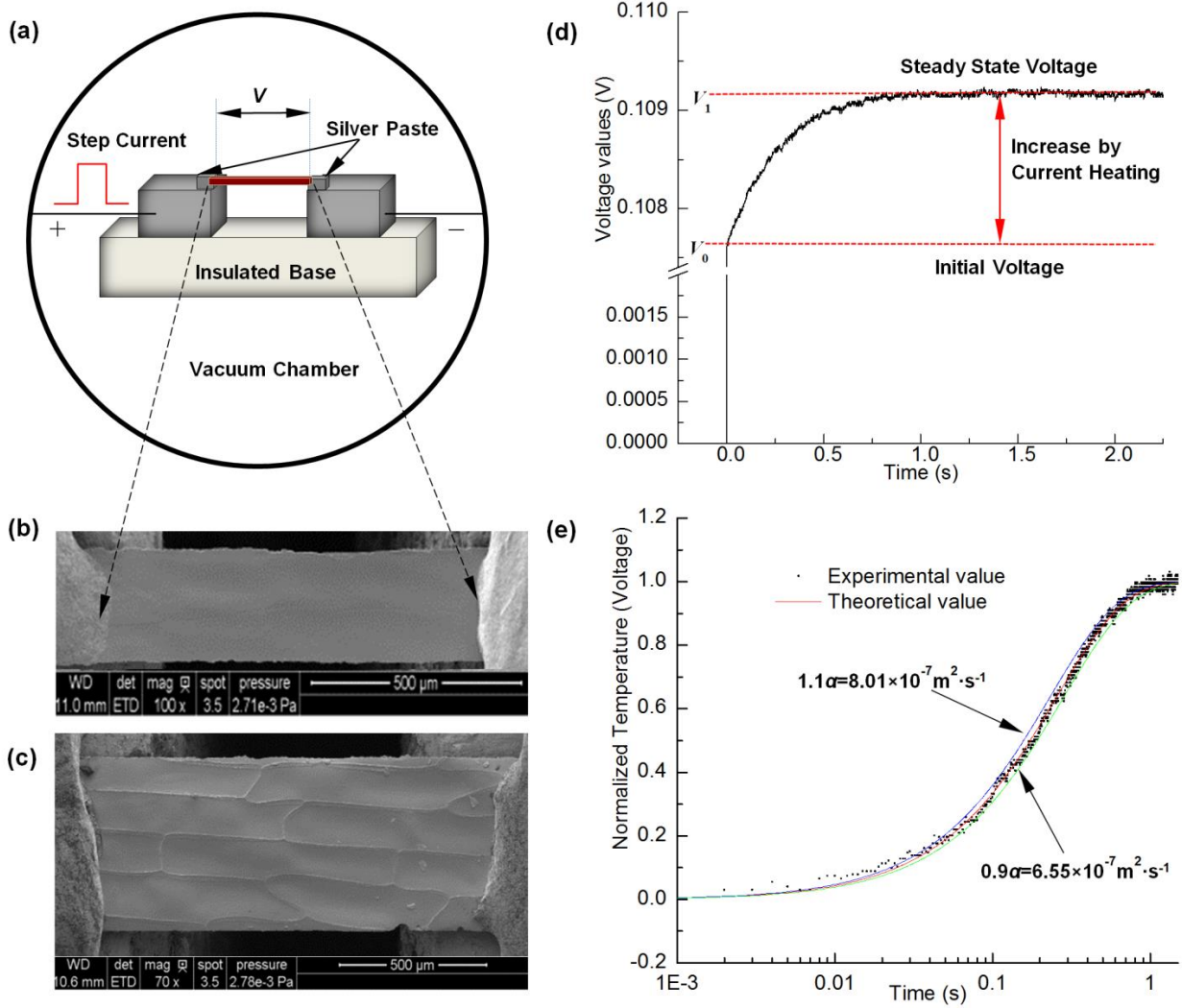

**Fig. 3** (a) Schematic of the experimental principle of the TET technique to characterize the thermal diffusivity of cell-scale lignocellulose. (b-c) SEM graphs of a sample connected between two electrodes. From figure (c) we can clearly see the cell is connected between the two electrodes in the cell length direction. (d) A typical *V-t* profile recorded by the oscilloscope for Sample 1 induced by the step DC current. (e) TET fitting results for mono-layer cells (Sample 1). The figure consists of the normalized experimental temperature rise versus time, theoretical fitting results, and another two fitting curves with ±10% variation of $\alpha_{eff}$ to demonstrate the uncertainty of the fitting process (blue line is for +10%, and green one is for -10%).

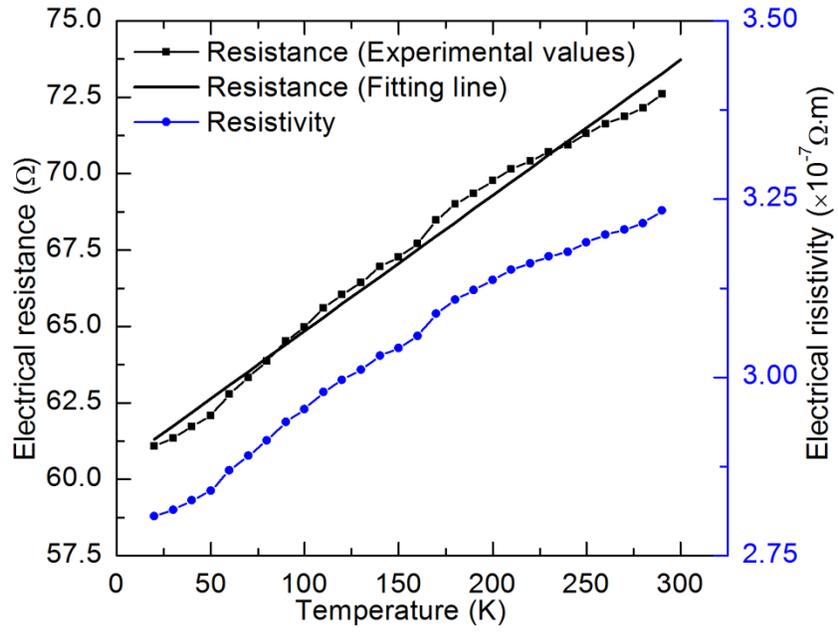

**Fig. 4** Temperature dependence of the electrical resistance and resistivity of Sample 2.

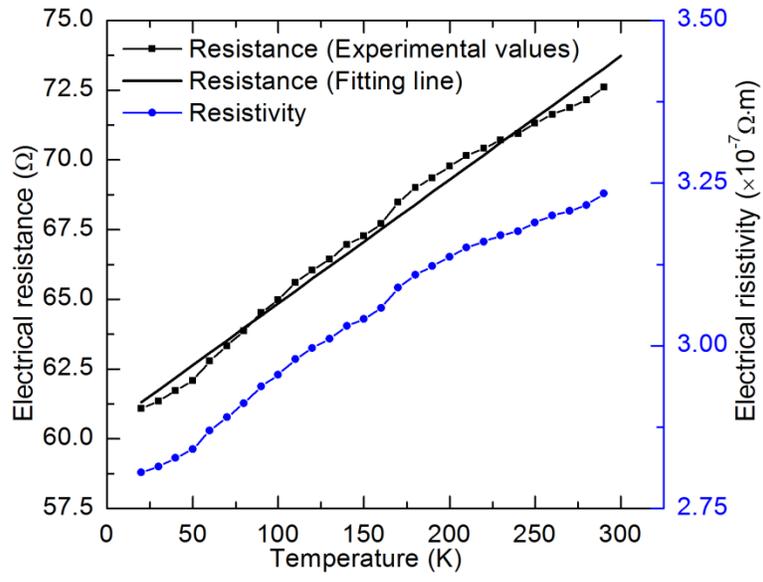

**Fig. 5** Temperature dependence of the cell-scale lignocellulose's volumetric specific heat.

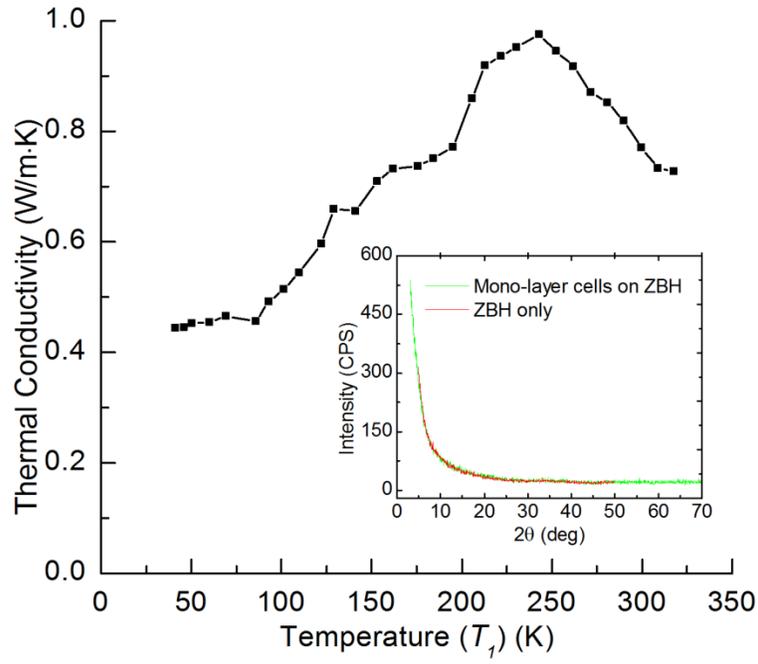

**Fig. 6** Temperature dependence of cell-scale lignocellulose's thermal conductivity. The inset shows the x-Ray Diffraction results of cell-scale lignocellulose.

## GRAPHICAL ABSTRACT

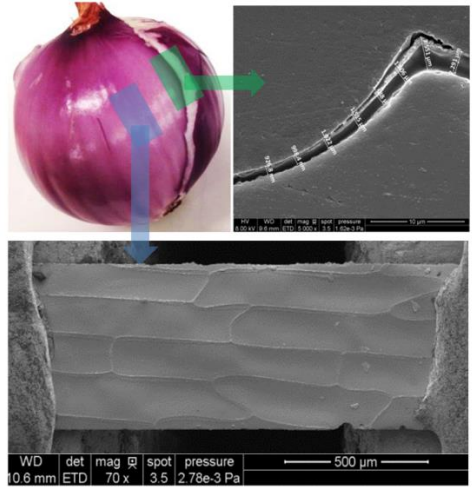